\documentclass[preprint]{emulateapj} 

\usepackage{natbib}
\usepackage{hyperref}
\hypersetup{colorlinks,citecolor=Blue,linkcolor=Red,urlcolor=Blue}
\usepackage{amsmath}
\usepackage{empheq}
\usepackage[usenames,dvipsnames]{color}
\allowdisplaybreaks

\begin{document}
 
\title{Observational bias and the clustering of
distant eccentric Kuiper belt objects}
\author{Michael E. Brown}

\affil{Division of Geological and Planetary Sciences, California Institute of Technology, Pasadena, CA 91125} 
\email{mbrown@caltech.edu}

\begin{abstract}
The hypothesis that a massive Planet Nine exists in the outer
solar system on a distant eccentric orbit was inspired by observations
showing that the objects with the most distant eccentric orbits
 in the Kuiper belt
have orbits which are physically aligned, that is, they are
 clustered in longitude of perihelion and have similar orbital planes. 
Questions have remained,
however, about the effects of observational bias on these observations,
particularly on the longitudes of perihelion.
Specifically, distant eccentric Kuiper belt objects tend to be faint
and only observable near their perihelia, suggesting that the longitudes
of perihelion of the known distant objects could be strongly biased by
the limited number of 
locations in the sky where deep surveys have been carried out. 
We have developed a method to
rigorously estimate the longitude of perihelion bias for Kuiper
belt observations. We find that the probability that the 10 known
Kuiper belt objects with semimajor axis beyond 230 AU are
drawn from a population with uniform longitude of perihelion is 1.2\%.
Combined with the observation that the orbital poles of these
object are also clustered, the overall probability of detecting
these two independent clusterings in a randomly distributed
 sample is 0.025\%.
While observational bias is clearly present in these observations, it
is unlikely to explain the observed alignment of the distant eccentric
Kuiper belt objects.
\end{abstract}

\section{Introduction}
Recently, \citet[hereafter BB16]{2016AJ....151...22B}
 demonstrated that the most 
distant known objects in the Kuiper belt have orbits which are 
physically aligned, that is, they are clustered in longitude of perihelion.  
Such an alignment is unexpected, as differential 
precession will destroy any such alignment on a 10-100 Myr time scale. 
BB16 demonstrated that a distant giant planet in an eccentric orbit --
referred to here as Planet Nine -- could maintain an alignment
for the age of the solar system if longitude of perihelion
of the eccentric orbit of Planet
Nine is oriented 180 degrees away from those of the clustered 
Kuiper belt objects
(KBOs). To date, no other viable explanation for a physical alignment of
the orbits of distant eccentric KBOs has been proposed.

Previously, \citet[hereafter TS14]{2014Natur.507..471T} 
had noted that distant KBOs were clustered in
{\it argument} of perihelion, $\omega$, a parameter which corresponds not to
physical alignment but to a specific 
internal-orientation-with-respect-to-itself 
of an orbit. 
While TS14
speculated that a distant planet might be responsible, no
mechanism for clustering $\omega$ of a population of
KBOs by a planetary 
perturber without also having a physical longitude of perihelion clustering
has been found. \citet[hereafter MM16]{2016MNRAS.457L..89M} 
instead demonstrated that a massive disk of
eccentric KBOs will generate an inclination instability which
will naturally lead to clustering in $\omega$. To date,
no other viable explanation for a clustering of $\omega$
(that does not also include clustering of longitude of perihelion)
of distant eccentric KBOs has been proposed.

While the $\omega$ clustering reported by TS14 is
robust and cannot be caused by any observational bias (see below), 
the longitude of
perihelion clustering reported by BB16 is certainly 
subject to observational bias. As a simple example, a magnitude
limited survey will preferentially find objects which are
near their perihelion position where they are brightest, and if such
surveys are biased in the longitudes at which they are carried out,
that bias will be reflected in the longitude of perihelion
distribution found. Given that biases in survey longitudes are known to 
exist (mainly, but not exclusively, caused by avoidance of
the galactic plane), the possibility of a bias in measured 
longitude of perihelion should be carefully considered. BB16
made a simple argument that the most distant eccentric KBOs should
not be significantly more biased than not-quite-as-distant eccentric
KBOs -- which show an essentially uniform distribution of longitude of 
perihelion --
but it is clear that the bias towards finding objects at perihelion
grows with eccentricity, so it is not obvious how applicable this
simple argument is, particularly for the most eccentric orbits.

Because of this potential uncertainty about observational bias, the 
speculation that
the longitude of perihelion clustering might be purely an observational
selection effect has been suggested
\citep{2016AJ....152..221S,2017AJ....153...63S, 
2017AJ....153...33L,2017arXiv170401952B}.
Assessing the impact of observational bias in longitude of perihelion
is critical to understanding whether the observations point to a 
self-gravitating massive outer disk or to the presence of a giant ninth 
planet. Here we develop a rigorous method to estimate the longitude of
perihelion bias for distant eccentric KBOs. We apply the method to
the distant eccentric KBOs originally identified by
BB16 and to those that have been identified since to assess
the possibility of the presence of Planet Nine.

\section{Significance of $\omega$ clustering}
Before discussing biases in the longitude of perihelion, we
quickly discuss the argument of perihelion, $\omega$, and
show how the observed clustering around $\omega\sim 0$
cannot be caused by observational 
bias, even though
there {\it are} clear observational biases in
$\omega$ for eccentric objects. In particular, eccentric
objects with $\omega$ near 0 or 180 degrees come to perihelion 
and are thus brightest around the heavily observed ecliptic,
so one would expect eccentric objects to be found preferentially
around $\omega=0$ and 180 degrees even for a uniformly distributed
population. 
At the moment of discovery, however, an object with
$\omega=\omega\prime$ and one with $\omega=180-\omega\prime$ differ only
in the direction of the ecliptic latitudinal component of their 
velocity vectors. 
As pointed out by TS14,
there is no possible way to design a survey to be biased
in favor of finding objects close to $\omega= 0$ at the expense
of objects with $\omega$ close to 180 (or vice versa), yet the
distant eccentric KBOs show this effect strongly.

While calculating the full observational selection bias of $\omega$
is not possible, it is trivial to calculate the probability that 
objects would be exclusively clustered around $\omega= 0$ or around 180.
In the original analysis, TS14 found that the 12 most distant 
eccentric KBOs -- those with semimajor axis 150 AU and greater --
cluster within 43 degrees of $\omega=0$. The 
probability that 12 such objects would cluster around either 0 or 180
is simply $2\times 2^{-12}$, or 0.04\%. Note that here and throughout
this paper we refer to KBOs as all multi-opposition solar system objects with
perihelion distance beyond Neptune's orbit.

Since the original work of TS14, 9 new KBOs with semimajor axis
150 AU or greater have been found. Of these, 7 are closer to $\omega=0$ than
to $\omega=180.$ The probability that 19 or more of 21 objects would be
so clustered is $2\times 2^{-21} \times [C(21,2)+C(21,1)+1]$ where
C(n,m) is the number of independent combinations of $m$ objects
from a population of $n$. The probability of this occurrence is thus just
0.022\%.

No sophisticated debiasing needs to be done to show that the 
clustering in $\omega$ is highly significant. This strong
signal -- unexplained by the mechanism proposed by TS14 --
led MM16 to the realization that this clustering could be
caused by a massive distant disk causing an inclination
instability in the outer solar system. 
A distant eccentric Planet Nine, in contrast, clusters longitude of
perihelion and pole position, rather than $\omega$
\citep{BroBat17}. 
A population of
longitude-of-perihelion-aligned orbits with poles clustered 
around a position offset from the
north ecliptic pole will generally, but not exclusively, also
have clustered $\omega$. Unfortunately no simple calculation
gives pole position bias, so we continue to use the clustering
in $\omega$ as an imperfect statistical proxy for clustering in pole position.

Clustering of distant eccentric KBOs
in $\omega$ (or, alternatively, in pole position) is firmly established.
To date, the only viable explanations for this clustering
is either the inclination instability proposal of MM16 
or the Planet Nine proposal
of BB16. These proposals differ most 
in their predicted distribution
of longitude of perihelion. The inclination instability shows no preference
for clustering in longitude of perihelion, while Planet Nine confines the
longitudes. 

If in fact there is no longitude of perihelion clustering, the robust
clustering in $\omega$ is
currently only explainable by the presence of a massive outer disk of
material inducing a inclination instability through self-gravity,
as proposed by MM16. If, on the other hand, the clustering in longitude
of perihelion is a true effect, rather than an apparent one caused by
observational bias, Planet Nine remains the only currently proposed
explanation.  We now examine observational biases in longitude of 
perihelion to determine which of these hypotheses appears more likely.

\section{Observational bias in longitude of perihelion}
The best method for determining the effects of observational bias on
the known KBOs would be to have complete information of all of the
surveys conducted to date, including their depth, precise coverage,
and their efficiency. Such information is unknown for the majority
of the surveys that led to the discoveries of the cataloged KBOs. In
many cases, nothing is published about the discovery survey;
the existence of the object is simply cataloged by the IAU Minor 
Planet Center (see http://www.minorplanetcenter.net/iau/Unusual.html).
The cataloged information is sufficient 
to determine the ecliptic longitude, ecliptic latitude, heliocentric
distant,
and brightness of every object at the time of its discovery.

We develop a novel method to use
the discovery circumstances of
the ensemble of all KBOs in the catalog to rigorously estimate the 
statistical distribution
of longitudes of perihelion expected for distant eccentric KBOs.
Conceptually, the method relies on the idea that each KBO discovery
can be thought of as a survey that could have discovered a distant
eccentric KBO had that object been bright enough and in the same place.
We proceed as follows:
for each distant eccentric KBO (a "parent object") we 
construct a synthetic population of new objects assuming
an identical absolute magnitude and identical
orbital elements for a 
uniformly selected longitude of perihelion and mean anomaly
(in practice we also assume
symmetry across the ecliptic plane, so our constructed population
also includes orbits where we replace $\omega$ with $-\omega$). 
We tally the ecliptic longitude, ecliptic latitude,
heliocentric distance, and expected magnitude of each object
in the synthetic population.
We call this the "uniform population" of the parent KBO. 
Next, for every KBO discovery in the catalog
we assess whether or
not one
of the uniform population of the parent object 
exists at the ecliptic longitude and latitude of the
observation and if that member of the uniform population
is brighter than the actual detected KBO. If so, we
known that a survey was being undertaken at that point
that could have detected one of the uniform population.
(In practice, we look for members of the uniform
population within 1 degree of the discovery location
of the KBO discovery.)
Finally,
we tabulate the longitudes of perihelion of 
the members of the uniform population that could have 
been detected at that discovery location.
We now know that the survey that resulted in that particular
KBO discovery was
sensitive to members of
our uniform population if they had had a particular longitude of perihelion. 
This
procedure is repeated for every cataloged KBO discovery to determine the 
probability distribution function of the longitude of perihelion
of the parent KBO
assuming that the population
is uniformly distributed in longitude of perihelion and mean anomaly.

A concrete example makes this procedure more clear. Consider 2013 RF98,
the most eccentric of the objects originally identified by BB16, as
the parent object of a uniform population.
Next, consider the discovery
of a randomly cataloged KBO, 2015 GP50, which, at the moment of 
discovery, had an
ecliptic latitude of -11.2 degrees and a magnitude of 24.8. Examining
the orbit of 2013 RF98, we find that it crosses -11.2 degrees twice,
once 27 degrees from perihelion, when it has a magnitude of 24.6, and once
closer to aphelion, when it has a magnitude of 28.7.
Near aphelion, the uniform population would not have been detectable
at this latitude, but at its magnitude closer to perihelion it
could have been detected by the observation that discovered 2015 GP50.
The KBO 2015 GP50 was discovered at a longitude of 196 degrees, thus
the member of the uniform population that is detectable has a longitude of
perihelion of $196-27$=168 degrees. This specific observation is thus
biased to finding this specific longitude of perihelion for this
specific parent object.
If we now consider not
a single KBO discovery, but {\it all} KBO discoveries, we find a 
statistical distribution of the longitudes of perihelion in which 
discoveries of the 2013 RF98 uniform population could have been made. We
thus create a separate statistical distribution of expected
longitudes of perihelion for each distant eccentric KBO.

This conceptual framework relies on the assumption that
KBO discoveries roughly represent the coverage 
and depth of the combined surveys. This assumption is clearly
false for the  latitude distribution, where more KBOs are discovered
at low ecliptic latitudes simply because of their greater numbers.
We correct this bias
by scaling by expected density of KBOs at a given latitude.
To approximate this expected density we use the method developed by
\citet{2001AJ....121.2804B} to determine the
 inclination distribution and convert
it to a latitudinal distribution assuming circular orbits. The final 
results are not sensitive to the precise latitudinal distribution chosen.

A second way in which the assumption that KBO discoveries are uniform
with search area
is violated is in the known longitudinal
bias in the discovery of resonant KBOs, which are over-discovered 
near their perihelion positions, which are related to the position
of Neptune. The easiest way to avoid this problem is to discard 
all discoveries of Plutinos, which are, by far, the most numerically prominent 
and most spatially correlated of the resonant objects. In practice we
simply discard all discovered objects with semimajor axes below 
40 AU. This constraint also forces us to only retain multi-opposition
KBOs with orbits known accurately enough to calculate this parameter.
We note, however, that relaxing this assumption makes the final
results of this analysis {\it more} significant. Nonetheless, we
conservatively retain this constraint.

One other important assumption is that a distant eccentric KBO could
always be discovered if a closer but fainter KBO was discovered at
its predicted location. This assumption can be violated if the discovery
survey is not sensitive to distant objects. Some of the nearest KBOs
and many Centaurs, for example, have been discovered in surveys searching
for near-earth objects, which do not have observational baselines
sufficiently long to be sensitive to more slowly moving distant objects. To
exclude these surveys we will only consider discoveries of objects at
distances greater than 30 AU. Even for normal KBO surveys, some of
the distant eccentric KBOs might not be detectable due to their low
rate of motion even though they are still brighter than the magnitude
limit. Sedna, for example, would be visible to a distance of 225 AU to
a survey that went to a depth of 25th magnitude, but few surveys are
sensitive to the slow motions of
such distant objects. We will thus place an upper bound
of 90 AU on the most distant object that any survey could see. This value
is probably a conservative estimate and will have the effect of making
longitude of perihelion biases stronger than they might be in real life.
In total, we use the observations of 1248 objects to determine our
expected distributions.

Using all of these constraints, we calculate expected statistical
distributions of longitude of perihelion for each of the 10 known KBOs
with semimajor axis beyond 230 AU. These include the six originally 
identified by BB16 and the four that have been discovered since that time
(Figure 1).  The expected longitude of perihelion distributions are highly
structured and highly individual. One trend is easily seen. The brightest
objects (Sedna, 2007 TG422) are among the most uniform in their expected 
discovery distributions. Many surveys could have found the very bright
Sedna even quite far from its perihelion, for example. The structure
seen in the remaining objects is only understandable after analysis.
The distributions of 2012 VP113 and 2013 RF98, for example, are the
most non-uniform. These two objects both come to perihelion at 
high ecliptic latitude, so their populations are primarily observable 
at high latitude, yet few surveys reach the required depth at these
latitudes. The longitudes of perihelion of these objects are highly
biased by the limited number of surveys that could have detected
such a population. The structures in the other distributions are
similarly functions of perihelion latitude, brightness, and the 
distribution of surveys in the sky.
\begin{figure}
\plotone{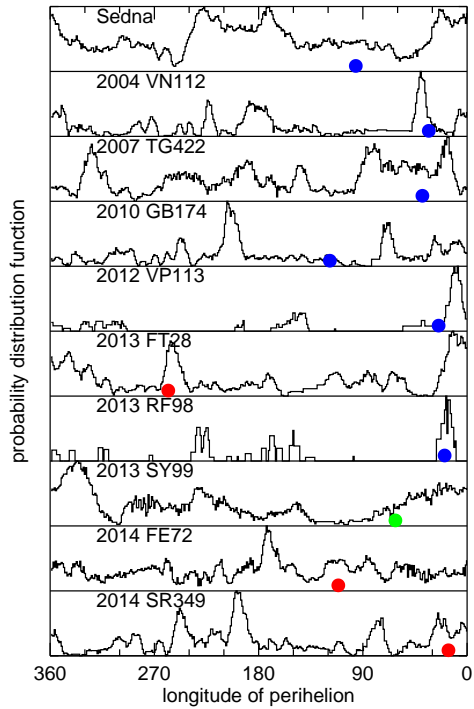}
\caption{Calculated probability distribution functions for the expected
distribution of longitude of perihelion assuming a population of
objects with identical orbital elements but uniformly distributed in
longitude of perihelion and mean anomaly. The colored dot shows the 
actual longitude of perihelion of each object. The blue dots
note the 6 KBOs originally discussed by BB16, the red dots
show the newer discoveries of \citet{2016AJ....152..221S}, while the green
dot shows the discovery of \citet{2017arXiv170401952B}.
While observational
biases are strong in the expected distribution of longitude of
perihelion, it is clear that for nearly all objects the bias towards
discovering the object with its actual longitude of perihelion 
is not severe.}
\end{figure}

\section{Comparison to observations}
We now see that measurement of the longitude of perihelion for 
a distant eccentric population is highly biased by the specific 
area of the sky and depth of individual surveys, as expected. 
With our determination of this bias, we can now examine whether 
the discoveries of distant eccentric KBOs are consistent
with being selected from a distribution which is uniform in 
longitude of perihelion 
or if, indeed, they are clustered. 

We first consider the six distant KBOs which BB16 reported 
as clustered in longitude of perihelion: Sedna, 2004 VN112, 2007 TG422, 2010 GB174, 2012 VP113, and 2013 RF98. At the time, these were all of the
known KBOs with semimajor axis beyond 230 AU.
 To understand the statistically expected
distribution of longitudes of perihelion for these bodies assuming
a uniformly distributed population, we perform 100,000 population 
samplings in which we create a new selection of 6 detected KBOs by
randomly choosing a longitude of perihelion
from the expected probability density function for each of
the 6 objects. We then examine the statistics of these 100,000 realizations.

The longitudes of perihelion of the 6 real objects are distributed such
that the maximum angle between any pair of angularly adjacent
objects is 260.9 degrees (Figure 2).
For the 100,000 realizations of this population assuming a uniform
distribution in longitude of perihelion, the maximum angle between
two angularly adjacent objects is 260.9 degrees or higher in only 1437 cases. If
the longitudes of perihelia of distant eccentric objects are uniformly
distributed, we would expect a longitude clustering as tight
as the one observed only 1.4\% of the time. 
We also compute the Rayleigh $z$ statistic of the data and the random sample and
find that Rayleigh $z$ value of 0.80 of the data is exceeded only 2.0\% of the time
in the random data.  In contrast, the simple estimate
from BB16 (which also took into account the clustering in 
perihelion latitude which is ignored here)
suggested that the clustering should only be observed
0.7\% of the time. We regard the rough agreement from 
independent ways of estimating the significance of this result as encouraging.
\begin{figure}
\plotone{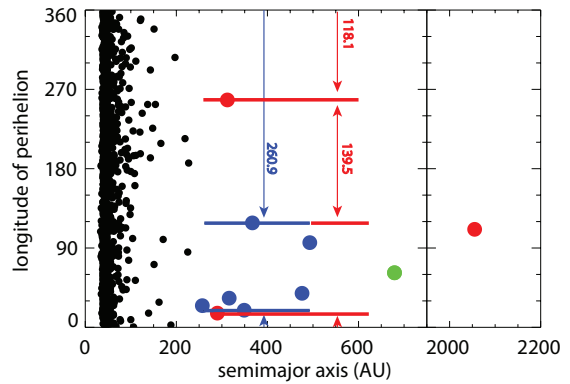}
\caption{Longitude of perihelion of KBOs as a function of semimajor axis
(note the axis change to include 2014FE72 with a semimajor axis of 2055 AU).
The blue objects are the original six discussed by BB16, which included
all KBOs with semimajor axes greater than 230 AU known at the time.
The red points are subsequent discoveries from \citet{2016AJ....152..221S}
 while the single
green point is from \citet{2017arXiv170401952B}. Angular distances
between subsets of KBOs discussed in the text are noted.}
\end{figure}

With only the six objects defined here, the probability of $\omega$
clustering becomes $2\times 2^{-6}$ or 3.1\%. Restating these
two findings, we see that there is only a 3.1\% chance that the
values for $\omega$ are equally distributed about 0 and 180 degrees,
suggesting that some mechanism is clustering the distant KBOs in $\omega.$
We likewise find that there is only a 1.4\% chance that the longitudes
of perihelia are distributed uniformly, suggesting that these
values are likewise clustered. The combined probability that both of
these clusters would be found in random data is thus 0.043\%.

Since the original analysis of BB16, \citet{2016ApJ...824L..23B} demonstrated
that the Planet Nine hypothesis, in addition to clustering distant 
eccentric KBOs in longitudes of perihelion opposite to that
of Planet Nine (the ''anti-aligned population''), a smaller population
of objects with longitudes of perihelion {\it aligned} with Planet Nine
should also exist (the ''aligned population''). 
Four new distant eccentric KBOs (with semimajor axis greater than 230 AU)
have been discovered since the initial analysis \citep{2016AJ....152..221S,2017arXiv170401952B}. 
Of these, three fit well with the anti-aligned population (2013 FT28,
2014 FE72, and 2013 SY99), while one is consistent with being the
first recognized member of the aligned population (2014 SR349).
The realization that we now expect two separate oppositely-oriented
populations requires a different metric for assessing the match
between the expected and observed population. Instead of examining
the largest angle between the longitudes of perihelion of
any two angularly adjacent objects, we look at
the second largest angle between any two angularly adjacent objects. 
In a population with
two well separated oppositely oriented groups, 
this second largest angle will be maximized. 
In practice, we would have also considered 
the observed population of objects clustered 
if there had only been a single cluster rather than two. Such 
a cluster would have a large largest angle but a small second largest
angle. To overcome this problem we take either the second largest
angle or half of the largest angle, effectively mimicking the effects
of two populations even if only one is observed.
In the real population of 10 distant eccentric KBOs, the separation
in longitude of perihelion
between the anti-aligned group and 2013 SR349 is 139.5 degrees
in one direction and 118.1 in the other direction. Our second largest
angle is thus 118.1 degrees. We again perform 100,000 random iterations
and compare this expected population to the real observations.
In only 1201 cases is the second-largest separation between longitudes
of perihelion as large or larger than 118.1 degrees. The probability
that the distant eccentric KBOs
 would be distributed in longitude of perihelion
as extremely as the observations are if the underlying distribution
were uniform is 1.0\%.  The calculated Rayleigh $z$ statistic of
this population (which is only sensitive to a unimodal distribution)
with a value of 0.62 is exceeded in 2.2\% of the random 
sample. While the discovery of four new distant eccentric
KBOs might have been expected to increase our confidence in these
populations significantly,
the realization that we are observing two opposing rather than one
single clustered population necessarily dilutes the signal.

One of these objects (2014 FT28) has $\omega=286$ degrees, while the
rest are clustered about zero. \citet{2016AJ....152..221S} use this fact to exclude 
2014 FT28 from consideration as part of the clustered population,
however
\citet{BroBat17} show that such objects
are indeed expected in the Planet Nine hypothesis. They are simply
precessing around a displaced pole but have temporarily circulated past
the north ecliptic pole. Ecliptic-based Keplerian orbital elements
are a poor descriptor of the orbits in this case. Nonetheless,
we continue to use the simple measure of $\omega$ 
as a proxy for pole clustering and find that if the true values of $\omega$
are uniform about 0 and 180, the probability that
9 or more of 10 values for $\omega$ would be clustered about either
is 2.1\%. The combined probability of both of these parameters being
clustered is 0.025\%.

The original $\omega$ clustering discussed by
TS13 which led to the MM16 hypothesis of inclination 
instability includes all KBOs with semimajor axis 150 AU and larger.
To date, 21 such KBOs are known, and 19 have $\omega$
closer to 0 than to 180 degrees. 
The probability of this
clustering occurring randomly is a mere 0.022\%. 
Examination of
the longitude of perihelion (Figure 2)
shows that this parameter, too, has some
structure down to a semimajor axis of 150 AU, but it is
clear that the clustering in longitude of perihelion is beginning
to break down. This behavior was seen in the population
simulations of \citet{BroBat17} where it was noted that the
longitude of perihelion merged from being highly clustered 
at large semimajor axis, to moderately to not-at-all clustered
as semimajor axis decreased. We thus do not expect the longitude
of perihelion cluster of objects beyond 150 AU to be as significant
as those beyond 230 AU that we initially considered.
We nonetheless assess the observational
biases.
Once again performing 100,000 iterations of a population
uniformly distributed in in longitude of perihelion we
find that 43590 have a second-largest longitude of
perihelion angle between two KBOs as large or larger than 48.2 
degrees, the value seen in the real data (while 39753 have
a largest angle of 67.9 or greater like the data). The
Rayleigh $z$ test likewise shows no significance to this clustering.
In short,
for objects with semimajor axis 150 AU and larger
there is no statistically significant clustering of longitude of
perihelion into one or two groups, yet the cluster of $\omega$
is highly significant. This discrepancy shows, we believe,
the expected blending of the high semimajor axis
longitudinally clustered objects into the lower semimajor axis
background population and
the uncertainty as to where precisely to draw the line for
distant objects which are and which are not affected by Planet Nine.
At greater semimajor axes, the longitude of perihelion clustering 
is robust as expected.

\section{Conclusions}

We have shown that measurement of the longitude
of perihelion of a population of distant eccentric KBOs is
subject to considerable observational bias, but that this 
bias is unlikely to be responsible for the observed
clustering in longitude of perihelion. 
If distant eccentric KBOs were uniformly distributed in 
longitude of perihelion, observations of
the original six objects with semimajor axis beyond 230 AU
which led BB16 to suggest the existence
of Planet Nine would only find the extreme clustering observed
1.4\% of the time. Including the four most recently discovered
KBOs with semimajor axes beyond 230 AU drops that probability to
1.2\%.

Clustering in $\omega$ of distant eccentric KBOs
is firmly established. Thus
determination of the veracity of the clustering in longitude of
perihelion is critical to understanding the gravitational forces
sculpting the outer solar system. 
With no longitude of perihelion clustering,
the only currently proposed viable mechanism for causing $\omega$ clustering is 
the MM16 suggestion
of a massive outer disk causing an inclination instability 
and $\omega$ clustering. If, on the other hand, 
longitude of perihelion and pole position (which roughly manifests
itself as $\omega$ for an offset pole) are clustered,
Planet Nine is the only currently proposed viable hypothesis. By 
rigorously estimating the effects of observational bias, we have shown here
that the Planet Nine hypothesis is by far the more likely of these scenarios.
The probability that the combination of the
 alignment of the longitudes of perihelion with the clustering
in pole position (using the $\omega$ proxy) that is seen in the KBOs
with semimajor axes beyond 230 AU would occur by chance
in a uniformly distributed population is only 0.025\%. While explanations
other than Planet Nine might one day be found to explain 
these observations, the significance of the observations themselves
appears secure.

\acknowledgements{We would like to thank Ann-Marie Madigan for the discussion
which inspired this analysis and Elizabeth Bailey, Konstantin Batygin, and 
Ian Wong for critical readings of the manuscript.}

\end{document}